%Paper: gr-qc/9412017
%From: menotti@ibmth.difi.unipi.it (Pietro MENOTTI)
%Date: Tue, 6 Dec 1994 10:36:00 +0100 (NFT)

\magnification=\magstep1
\hsize=16truecm
\vsize=22truecm
\baselineskip=20pt plus 1pt
\voffset=1truecm
\hoffset=0truecm
\parindent=0cm
\def\pmb#1{\setbox0=\hbox{#1}%
  \kern-.025em\copy0\kern-\wd0
  \kern.05em\copy0\kern-\wd0
  \kern-.025em\raise.0433em\box0}
\def\bfxi{\pmb{$\xi$}}
\def\bs{\bigskip}

\def\tr{{\rm tr}}

\def\Star{1}
\def\Vile{2}
\def\Callum{3}
\def\Wein{4}
\def\DJtH{5}
\def\NelReg{6}
\def\AchTow{7}
\def\Witten{8}
\def\Scho{9}
\def\MSener{10}
\def\HawEll{11}
\def\Gott{12}
\def\Cut{13} % Cutler, Ori, Kabat, DJt
\def\CFGO{14}
\def\Moncrief{15}
\def\FerWal{16}
\def\ModTol{17}
\def\MS{18}
\def\Goedel{19}
\def\Tipler{20} % also Hawking
\def\tHuno{21}
\def\tHdue{22}
\def\MSctc{23}
\def\GerHor{24}
\def\ADM{25}
\def\HosNak{26}
\def\NelRegdue{27}
\def\Carlip{28}
\def\Carlipdue{29}
\def\CarNel{30}
\def\Wael{31}

\hfill IFUP-TH-70/94
\vskip 5cm
\centerline{\bf Gravity in 2+1 dimensions}

\bs
\centerline{ Pietro Menotti}

\bs
\centerline{Dipartimento di Fisica della Universit\`a di Pisa}
\centerline{ and}
\centerline{ INFN Sezione di Pisa}
\bs
\bs

\centerline{\bf Abstract}

A review is given of some classical and quantum aspects of 2+1
dimensional gravity.

\bs
\bs
\vskip 4cm
Two lectures given at the XI Italian Relativity Meeting, Trieste 26-30
September 1994.

\bs
Work supported in part by M.U.R.S.T.
\vfill
\eject

{\bf 1. Introduction}

Gravity in $2+1$ dimensions [\Star] can be given two different
meanings: one as
the study of special solutions of $3+1$ dimensional gravity in
presence of a space-like Killing vector; the other as a simplified
model of gravity which can lend itself as playground to hopefully learn
how to deal with problems in 3+1 which up to now have defeated our
comprehension. The first aspect is of interest in connection of
cosmic strings.
%It is worth mentioning that not all solutions with an
%open space like Killing vector in $3+1$ dimensions give rise to
%solutions of $2+1$ dimensional gravity (Tipler, Kramers); on the other
%hand all solution of $2+1$ dimensional gravity have a counterpart in
%$3+1$ dimensions.

One starts by counting the number of independent components
of the Riemann tensor as a function of the space-time dimensions $D$.

%We shall use the conventions
%$$
%R^\mu_{~\nu\mu\rho}=R_{\nu\rho}= {\rm Ricci~tensor}
%$$
%$$
%R^\mu_{~\mu}=R={\rm scalar~ curvature}
%$$
In $D=2$ due to the
symmetries of the Riemann tensor we have only one independent
component $R_{0101}$ and actually we can write
\def\unouno{1.1}
$$
R^{\mu\nu}_{~~\lambda\rho}=(\delta^\mu_\lambda \delta^\nu_\rho-
\delta^\mu_\rho \delta^\nu_\lambda) R/6.\eqno(\unouno)
$$
In $D=3$ due to the antisymmetry of $R_{\mu\nu\lambda\rho}$ in the pairs
$(\mu\nu)$ and $(\lambda\rho)$ we have that each pair can assume only three
values and due to the symmetry under the exchange of the two pairs we have that
the Riemann tensor has only $6$ independent values i.e. as many as the number
of independent components of the Ricci tensor.
We can also write
\def\unodue{1.2}
$$
R^{\mu\nu}_{~~\lambda\rho}=\epsilon^{\mu\nu\kappa}\epsilon_{\lambda\rho\sigma}
( R^\sigma_\kappa - \delta^\sigma_\kappa {R\over
2})=\epsilon^{\mu\nu\kappa}\epsilon_{\lambda\rho\sigma}
G^\sigma_\kappa\eqno(\unodue)
$$
being $G^\sigma_\kappa$ the Einstein tensor.

Einstein's equations are written as
\def\einstein{1.3}
$$
R_{\mu\nu}-{g_{\mu\nu}\over 2} R={8\pi G \over c^4}T_{\mu\nu}
\eqno(\einstein)
$$
where $G$ is Newton's constant whose dimension is
$l^{D-3} m^{-1} c^2$. If we consider in $D=4$ a geometry in which the
$g_{\mu\nu}$ are independent of $x^3$ and $g_{\mu\nu}$ has the structure
\def\unoquattro{1.4}
$$
g_{03}=g_{30}=0,~~~~g_{33}=1
\eqno(\unoquattro)
$$
%$$
%g_{\mu\nu}=\pmatrix{* & * & * & 0 \cr
%                    * & * & * & 0 \cr
%                    * & * & * & 0 \cr
%                    0 & 0 & 0 & 1 }
%
%$$
%then the Ricci tensor assumes the form
%$$ R_{\mu\nu}=\pmatrix{* & * & * & 0 \cr
%                    * & * & * & 0 \cr
%                    * & * & * & 0 \cr
%                    0 & 0 & 0 & 0}
%$$
then $R_{0\mu}=R_{\mu 0}=0$
and for $\mu,\nu=0,1,2$ we have the three dimensional Einstein's
equations, where
$R_{\mu\nu}$ is just computed from the components $0,1,2$ of the four
dimensional metric $g_{\mu\nu}$.
{}From eq.(\einstein) we see that the components $3,\mu$ with $\mu=0,1,2$ of
the energy momentum tensor vanish, while  we have
\def\unosei{1.5}
$$
{8\pi G \over c^4}T_{33} = -{g_{33}\over 2} R\eqno(\unosei)
$$
which in general is not zero. The written formulas prove that to all
solution in 2+1 dimensions there corresponds a solution in $3+1$ with a space
like Killing vector and an energy momentum tensor which is simply related
to the one in $2+1$; in $3+1$ a tension develops along the third dimension.

The situation described above is interesting in connection of cosmic
strings [\Vile].

We note however that not all solutions of Einstein equations in $3+1$
dimensions with
an open space-like Killing vector give rise to solutions of $2+1$ dimensional
gravity. We can quote as an example the axial solutions given by
Levi-Civita [\Callum]
\def\unosette{1.6}
$$
ds^2=\rho^{-2m}[\rho^{2m^2}(d\rho^2+dz^2) + \rho^2 d\phi^2]-\rho^{2m}dt^2.
\eqno(\unosette)
$$
We see here that $g_{33}=\rho^{-2m}$ and as such not constant.

In all dimensions $T_{\mu\nu}=0$ implies $R_{\mu\nu}=0$ but in $2+1$
dimensions $R_{\mu\nu}=0$ implies $R_{\mu\nu\lambda\rho}=0$ and
thus where matter is absent space-time is flat. This is the most
important property of $2+1$ dimensional gravity.

\bs
{\bf 2. Energy, momentum and angular momentum in $2+1$ dimensions}

After these general comments we come to the definition of the
fundamental quantities in the theory, i.e. energy-momentum and angular
momentum.

The energy-momentum and angular momentum of a system in general
relativity in $D$
dimensions with $D\ge 4$ is defined by means of the flux of a
pseudotensor through a $D-2$ dimensional space-like surface located at
space infinity. For example for the energy-momentum we have in four
dimensions [\Wein]
\def\dueuno{2.1}
$$
{16\pi G\over c^4} {\cal P}_\lambda= -{1\over 2}
\oint\delta^{\sigma\alpha\beta}_
{\mu\nu\lambda}~g^{\nu\delta}~ \Gamma^\mu_{\delta\beta} ~ d
S_{\sigma\alpha}\eqno(\dueuno)
$$
where $dS_{\sigma\alpha}$ is the element of a space-like two dimensional
surface at space infinity.

%The holonomies of the three dimensional Poincar\'e group will play a
%fundamental role in all what follows. We consider the path ordered

It is well known that such energy-momentum can be
defined when the metric approaches at space infinity the minkowskian
metric.
%\cite{ADMLW}.
Such a definition cannot be taken over directly to $2+1$ dimensional
gravity; in fact it is true that the space outside the sources is
exactly flat, but the global structure of the space time at space
infinity is that of a cone with a non zero deficit angle.
%\cite{DJH}
We shall see e.g. that for a stationary, static distribution of matter such
a deficit angle equals the space integral of $T^0_0$
i.e. the total amount of mass of
the matter present in the system. The  result however does not hold
in general. With regard to  angular momentum the
situation is more complicated. One can consider the rather artificial
massless ( zero deficit angle ), and  then one can
carry over the usual procedure of $3+1$ dimensions and thus one is able
%\cite{DJH}
in the stationary case to identify the angular momentum with the time
jump which occurs in a synchronous reference system when one performs a
closed trip around the source [\DJtH].

A general definition of energy-momentum and angular momentum in
2+1 dimensions can be derived by computing the holonomies of the
$ISO(2,1)$ group, which were introduced in connection to the
formulation of pure 2+1 dimensional gravity as a Chern-Simon theory.
%\cite{Witten}.
%(The idea of defining mass and angular momentum  in
%2+1 dimensional gravity by means of holonomies is already contained
%in the  paper
%\cite{DJH}
%for a system of point-like spinless particles).

Lorentz and more generally Poincar\'e holonomies will play a fundamental role
in all what follows and thus we turn now to them.

Let us consider to start the Lorentz holonomy
\def\duedue{2.2}
$$
M={\rm Pexp} (-i \oint J_a\omega^a_\mu dx^\mu) \eqno(\duedue)
$$
where
\def\duetre{2.3}
$$
\omega^a_\mu={1\over 2}\epsilon^{a~c}_{~b}\omega^b_{~c\mu}\eqno(\duetre)
$$
and $J_a$ are the generators of the $SO(2,1)$ group with
commutation relations given by
\def\duequattro{2.4}
$$
[ J_a, J_b]=i \epsilon_{abc}J^c\eqno(\duequattro)
$$
and the traces given by
\def\duecinque{2.5}
$$
{\rm Tr}( J_a J_b)= -{{1}\over {2}}\eta_{ab}\eqno(\duecinque)
$$
where $\eta_{ab}\equiv {\rm diag}(-1,1,1)$. Here we shall consider for
definiteness the fundamental representation of SU(1,1).
It is not difficult to show that an element of $SU(1,1)$ can always be
written in the form $ \pm e^{-iJ_a\Theta^a}$ according to the
following cases:

%{\noindent\it Case 1.}
If $\Theta^a$ is a light-like vector
%In this case, the
%expansion
%(\ref{E2})
%
then in the fundamental representation we have
\def\duesei{2.6}
$$
M=\pm e^{-iJ_a\Theta^a}=\pm(I- i \Theta^a J_a).\eqno(\duesei)
$$
%all the remaining terms being equal to zero,  and taking the trace
%with  $J^b$ one obtains
%$$
%\Theta^b= \mp 2 i Tr( W J^b).\eqno(\duesette)
%$$

%{\noindent \it Case 2}.
If $\Theta^a$ is space-like
\def\dueotto{2.7}
$$
M=\pm e^{-iJ_a\Theta^a}=\pm (\cosh {\sqrt{\Theta^a \Theta_a}\over 2} I-
2 i  \displaystyle{{\sinh\sqrt{\Theta^a
      \Theta_a}/2\over \sqrt{\Theta^a \Theta_a}}}
(J_a \Theta^a)).\eqno(\dueotto)
$$
%Taking the trace  one finds
%$$
% Tr(J^b W)= \pm i \displaystyle{{\sinh{\sqrt{\Theta^a
%        \Theta_a}/2}\over \sqrt{\Theta^a \Theta_a}}} \Theta^b\eqno(\duenove)
%$$
%This determines completely $\Theta^b$,  as $\sinh$ is an odd monotonic
%function.

%{\noindent \it Case 3}.
If $\Theta^a$ is time-like
%The expansion in the
%fundamental representation is
we have
\def\duedieci{2.8}
$$
M=\cos {\sqrt{-\Theta^a \Theta_a}\over 2} I-
 2 i  \displaystyle{{\sin\sqrt{-\Theta^a
       \Theta_a}/2\over \sqrt{-\Theta^a \Theta_a}}}
(J_a \Theta^a).\eqno(\duedieci)
$$
%Taking the trace
%$$
%%\label{AAA}
% Tr(J^b W)= i \displaystyle{{\sin{\sqrt{-\Theta^a
%        \Theta_a}/2}\over \sqrt{-\Theta^a \Theta_a}}}
%\Theta^b.\eqno(\dueundici)
%$$
The $\pm$ alternative is a necessary one for $\Theta^a$ space-like or
light-like.

In the light-like and space-like case by computing $\tr M$ and $\tr (M
J^a)$ one determines completely $M$ and thus $\Theta^a$ while in the
time-like case $\Theta^a$ is completely determined in the range
$0\leq\sqrt{-\Theta^a\Theta_a} \leq 2\pi$.
%Eq.(\dueundici)
%(\ref{AAA})
%is not sufficient to determine $\Theta^a$, but combining it
%with the trace of $W$
%$$
%Tr(W)=2 \cos{\sqrt{-\Theta^a \Theta_a}\over 2}
%$$
%one finds that $\Theta^a$ is fixed if one  imposes that $\delta\equiv 8 \pi G
%M=\sqrt{-\Theta^a \Theta_a}$ satisfies
%$0\le \delta\le 2 \pi$ (open universe).
In this way we are also able to establish
whether  $\Theta^a\in V^+$ or $V^-$, where $V^+$ and $V^-$ denote the set
of  future and past directed time-like vectors.
%The use of the adjoint representation would
%have left us with the ambiguity ($\delta$ ,$\hat n$
%)$\leftrightarrow$ ($ 2 \pi -\delta$, $-\hat n$), where $\hat n$ is
%the versor  corresponding to the vector defined by
%(\ref{AAA}).
%In the
%construction discussed in the following section, however,
%this ambiguity is  avoided by continuity. On the
%other hand for a closed universe the total energy is defined  as the
%Euler number of the space 2-manifold.

In addition to  eq.(\duedue) one can consider the Poincar\'e holonomy given by
\def\duetredici{2.9}
$$
W={\rm Pexp}(-i \oint (J_a\omega^a_\mu +P_a e^a_\mu)dx^\mu)=
\pm e^{-iJ_a\Theta^a-i P_a E'^a}\eqno(\duetredici)
$$
which can also be written as
\def\duequattordici{2.10}
$$
W={\rm Pexp}(-i \oint P_a T^a_l e^l_\mu dx^\mu)\times {\rm Pexp}(-i \oint
J_a\omega^a_\mu dx^\mu)
= \pm e^{-iP_a E^a} e^{-iJ_a\Theta^a}
\eqno(\duequattordici)
$$
where
$$
E^a=\int_1^0 T^a_b(t)dt~ E'^b
$$
with
\def\duesedici{2.11}
$$
T^a_b(t)=[e^{-itJ^c\Theta^c}]^a_b \eqno(\duesedici)
$$
in the adjoint representation.
It is useful to represent $W$ as [\NelReg]
\def\duesedicix{2.12}
$$          W =\pmatrix{M & E \cr
                        0 & I   }
\eqno(\duesedicix)
$$
with multiplication rules
\def\duesediciy{2.13}
$$
\pmatrix{M_1 & E_1 \cr 0 & I   } \pmatrix{M_2 & E_2 \cr 0 & I   }=
\pmatrix{M_1 M_2  & {\cal M}_1 E_2+ E_1 \cr 0 & I   }
\eqno(\duesediciy)
$$
being ${\cal M}$ the transformation $M$ in the adjoint representation.

Let us now consider  the Poincar\'e holonomy for a closed loop; we can
perform a gauge transformation at the origin of the loop or
equivalently move the point
along the contour or move the whole loop in space provided it always
moves without intersecting matter. In all cases $W$ is subject to a
gauge transformation $W'= TWT^{-1}$
%$$
%W'=e^{-iP_a S^a} A W A^{-1} e^{iP_a S^a} = \pm
%e^{-iP_a E'^a} e^{-iJ_a\Theta'^a}
%
%$$
with
\def\duediciassette{2.14}
$$
M'=AMA^{-1}
\eqno(\duediciassette)
$$
where $A$ is an element of $SU(1,1)$  from which
\def\duediciotto{2.15}
$$
\Theta' ={\cal A}\Theta \eqno(\duediciotto)
$$
and
\def\duediciannove{2.16}
$$
E'={\cal A} E + (I-{\cal AMA}^{-1}) S\eqno(\duediciannove)
$$
being ${\cal M}$ and ${\cal A}$ the transformations $M$ and $A$ in the
adjoint representation and $S$ the translation vector of $T$.
So while $\Theta$ transforms like a Lorentz vector, $E^a$ transforms
inhomogeneously. We can envisage two invariants
$\Theta^a \Theta_a$ and $\Theta^a E_a$. The occurrence of such
invariants was noticed by Achucarro and Townsend [\AchTow] and by Witten
[\Witten ] and exploited by them to relate $2+1$ dimensional gravity
to Chern-Simon theory.

Thus from the Lorentz holonomy one can extract a vector under local Lorentz
transformations. Such a vector does not depend on deformations of
the loop keeping the origin fixed provided
the loop extends only outside the matter, and the square of the vector can be
naturally identified with the square of the mass of the system. Change
in the origin of the loop is equivalent to a Lorentz transformation
on the vector.

The energy momentum of the system
will be identified by ($c=1$ from now on)
\def\dueottobis{2.17}
$$
{\cal P}^a = \Theta^a /8\pi G
\eqno(\dueottobis)
$$
when $M$ is the Lorentz
holonomy computed along a closed contour which encloses all matter.

A major result of classical general relativity is the positive energy
theorem [\Scho] which states that the energy-momentum pseudovector as defined
in eq.(\dueuno) (something which we recall is possible in $D$ dimension
with $D\ge 4$ and for spaces which are asymptotically flat) is a
future directed time-like vector under transformations which are
asymptotically global Lorentz transformations. The ingredients for the
proof are the
energy condition and the existence of a space-like ``initial
condition'' $D-1-$dimensional hypersurface. A similar result can be
obtained in $2+1$ dimension [\MSener] with the definition of energy
given above eq.(\dueottobis)
but as we shall see now there are some differences.

We shall assume the existence of an edgeless space-like initial data
two dimensional surface $\Sigma$ with the topology of $R^2$. Let us
consider on $\Sigma$ a family of closed
contours  $x^\mu(s,\lambda)$ with  $0\le \lambda\le 1$,
$x^\mu(s,0)=x^\mu(s,1)$, and such that for   $s_1<s_2$ the contour
$x^\mu(s_1,\lambda)$ is completely contained in
$x^\mu(s_2,\lambda)$; moreover $x^\mu(s,\lambda)$ shrinks to a single
point for $s=0$.

It is easy to prove that the Wilson loop under the
deformation induced  by the parameter $s$ changes according to the
following  equation [\MSener]
\def\dueventi{2.18}
$$
{D M\over d s}(s,1)\equiv {d M \over d s}(s,1)+i [J_a \omega^a_{\mu}(s,1)
{d x^\mu\over d s}, M(s,1)]=
$$
$$
i M(s,1)\int^1_0 d\lambda M(s,\lambda)^{-1} R_{\mu\nu}(s,\lambda)
{d x^\mu\over d\lambda}{d x^\nu\over d s} M(s,\lambda)
\eqno(\dueventi)
$$
where $R_{\mu\nu}$ is the curvature  form in the fundamental
representation. Eq.(\dueventi) is true in any dimension but in 2+1
dimensions the curvature 2-form is given directly by the
energy-momentum tensor
\def\dueventuno{2.19}
$$
R_{\mu\nu}=8 \pi G ~\eta_{\mu\nu\rho} {\cal T}^{a\rho} J_a
\eqno(\dueventuno)
$$
with $\eta_{\mu\nu\rho}\equiv \sqrt{- g}~ \epsilon_{\mu\nu\rho}$ and
${ \cal T}^{a\rho} = e^a_\mu T^{\mu\rho}$.
%Substituting
%(\ref{Moto})
%into
%(\ref{Evol1})
%we have
%$$
%{D W\over d s}(s,1)= 8 \pi i G~ W(s,1)\int^1_0 d\lambda W(s,\lambda)^{-1}
%\eta_{\mu\nu\rho} T^{a\rho}J_a
%{d x^\mu\over d\lambda}{d x^\nu\over d s} W(s,\lambda)
%\eqno(\dueventidue)
%$$
Taking into account that
$-\displaystyle{\eta_{\mu\nu\rho}{dx^\mu\over d\lambda}{dx^\nu\over d
s}}$  is the area vector $N_\rho$ corresponding to the 2-dimensional surface
$\Sigma$, and thus time-like and the dominant energy condition (DEC)
[\HawEll] we have
\def\dueventitre{2.20}
$$
q^a(s,\lambda)\equiv {\cal T}^{a\rho} N_\rho\in V^+.\eqno(\dueventitre)
$$
%is time-like and future directed.
By substituting
%(\ref{q})
into eq.(\dueventi) we have
%(\ref{Evol2}) we have
\def\dueventiquattro{2.21}
$$
{D W\over d s}(s,1)=
%-8 \pi i G~ W(s,1)\int^1_0 d\lambda W(s,\lambda)^{-1}
%J_a q^{a}(s,\lambda)  W(s,\lambda)=
-8 \pi i G ~ W(s,1) J_a Q^a(s)
\eqno(\dueventiquattro)
$$
%where due to the similitude transformation $W^{-1} J_a q^a  W$, also
%the integrated
with $Q^a$ time-like and future directed.

If we  parameterize $M$ in the fundamental representation as
$M(s,1)=w(s) I - 2 i J_a \theta^a(s)$ the evolution equation gives
\def\dueventicinque{2.22}
$$
{d w(s)\over d s}= 4 \pi G  Q_a(s) \theta^a (s)\eqno(\dueventicinque)
$$
and
\def\dueventisei{2.23}
$$
%\label{evolP2}
{D \theta^b (s)\over d s}\equiv { d \theta^b(s)\over d
  s}+\omega^b_{c\mu}(s){dx^\mu\over d s} \theta^c (s)=4\pi G ~(w(s)
Q^b(s)+\epsilon^b_{~lm}\theta^l(s) Q^m(s)).
\eqno(\dueventisei)
$$

Such a system of differential equations can be discussed rigorously
[\MSener] with the following results:

For $s=0$ obviously $M(0,1)=w(0,1)=1$ and $\Theta^a=0$. Then as the
loop expands and starts including matter $\Theta^a$ becomes a
time-like future directed vector whose norm (i.e. the mass squared)
increases monotonically as the loop embraces more and more matter and
in the meantime $w(s)$ decreases monotonically from the initial value
$1$. The value $w(s)=-1$ can be reached in two ways, either with
$\theta^a=0$ or with $\theta^a$ light-like. In the first case the
universe has a time-like momentum and
$\sqrt{-\Theta^a\Theta_a}$ has reached the value $2\pi$; in absence of
further matter the
universe becomes a cylinder at large distances. Instead for $\theta^a$
light-like we have a light-like universe. If by still
expanding the loop we enclose more matter in case $1$ the momentum
stays time-like and the deficit
angle becomes larger that $2\pi$ and the universe closes kinematically
with the topology of a sphere; instead in case $2$, $\Theta^a$ becomes
space-like. An
example of such universe is the one envisaged by Gott [\Gott], built up by
two fast moving particles and which was subject to close scrutiny in
the literature [\Cut]. Again the rigorous discussion of the evolution
equations [\MSener] shows that it is not possible, by still adding matter to
such  space-like universe, to go back to an open time-like universe.

It is of interest to see what becomes of the energy for the stationary
static case. Here the dreibeins can be chosen
independent of time and of the form $e^a_{~0}= N \delta^a_0$,
$e^0_{~i}=0$ and we have
%\cite{DJH}
${\cal T}^{00}= N T^{00}$.
%where ${\cal T}^{00}$
%is the $00$ component of the energy-momentum tensor in the base given by
%the coordinates.
%Substituting into eq.
%(\ref{Evol1})
%we have
%$$
%{D W\over d s}(s,1)= 8 \pi i G~ W(s,1)\int^1_0 d\lambda W(s,\lambda)^{-1}
%\sqrt{-g} N T^{00}J_0
%\left({d x^1\over d\lambda}{d x^2\over d s}-
%{d x^2\over d\lambda}{d x^1\over d s}\right )
% W(s,\lambda)
%$$
%which is solved by
%$$
%W=V^{-1} (s)\exp(-8\pi iG J_0\int {\cal T}^0_0 \sqrt{-\gamma} dx^1 dx^2) V(s)
%$$
%where $\gamma$ is the determinant of the space metric and $V(s)$ is the
%holonomy of $\Gamma^a_s(s,1)$ computed along $x^\mu(s,1)$. Then we have
In this case the evolution equations can be solved explicitly
[\MSener] to get for the
mass of the system
\def\dueventisette{2.24}
$$
\delta=8\pi G m = 8\pi  G \int T^0_0 \sqrt{-\gamma} dx^1 dx^2=
4\pi   \int R^{(2)} \sqrt{-\gamma} dx^1 dx^2
\eqno(\dueventisette)
$$
where $T^0_0$ is the energy momentum tensor in the coordinate basis
and $\gamma$ is the determinant of the space metric,
in agreement with the result of [\DJtH]. The obtained result is expected
from special relativity taking into account that in $2+1$ dimensions
there are no gravitational forces, thus no gravitational potential
energy and thus static masses combine additively.
%\cite{DJH}.

Summing up in comparison to $D$ dimensions with $D\ge 4$ we have the
following differences:

\bs
1. In $3+1$ dimensions we can consider open universes of arbitrary
mass. In $2+1$ dimensions
%if one overcomes the limit of $2\pi$, either
%one closes the universe or one goes over to a space-like universe. In
%practice this is due to the fact that for $\delta <2\pi$ by adding
%matter, $\delta$ can only increase.
for small loops one always starts with an energy momentum
vector which is time-like future directed and such vector remains
time-like future directed putting together more matter until we reach
the $2\pi$ limit for the deficit angle. Then adding more matter one
goes over either to a
time-like closed universe or with proper
kinematical conditions one can reach a ``space-like'' universe in the
sense that the Lorentz holonomy becomes of the type $-e^{-iJ_a\Theta^a}$
with $\Theta^a$ space-like.

2. If the energy-momentum of the matter enclosed by a loop is
space-like in the sense mentioned above, then by enclosing more matter one
cannot go back to a time-like open universe. Thus we have a
generalization to all forms of matter satisfying the DEC of the theorem
by Carrol, Fahri, Guth and Olum [\CFGO]  which can be stated as follows: If a
subsystem of the universe has space-like momentum then either the universe
is closed or it has space-like momentum.

\bs
As we mentioned above $\Theta^a\Theta_a$ is not the only invariant of
the Poincar\'e holonomy; we have also $\Theta^a E_a$. This is
related to the angular momentum (which in $2+1$ dimensions has just
one component) by
\def\dueventotto{2.25}
$$
{\cal J}=-{\Theta_a E^a \over 8\pi G\sqrt{-\Theta_a\Theta^a}}
\eqno(\dueventotto)
$$
A first check of the correctness of our definition  comes from the
value it assumes for the ``Kerr'' solution in $2+1$ dimensions
%\cite{DJH}
\def\dueventinove{2.26}
$$
ds^2=-(d t + 4 G J d\theta)^2+(1-4 G m)^2 \rho^2 d \theta^2+ d\rho^2.
\eqno(\dueventinove)
$$
We find in fact that $\cal{J}$ of eq.(\dueventotto) coincides with the $J$
appearing in eq.(\dueventinove) which is related to the time shift $\Delta  t$
that
appears in a synchronous coordinate system when  one encircles once the
source $\Delta t= 4GJ\times 2\pi$. The $\cal{J}$ of eq.(\dueventotto) also
coincides (in the limit $m\rightarrow 0$) with the value obtained from
the $3+1$ dimensional
prescription in the case of a massless source with
angular momentum, for which the angular deficit at infinity is zero.

For the Poincar\'e holonomy one can write down evolution equations [\MSener]
similar to the ones written and discussed for the Lorentz
holonomy. These can be solved in the weak limit giving for $\cal{J}$
the expression of the angular momentum in special relativity and thus
providing further support to the identification (\dueventotto).
Thus formulas (\dueottobis, \dueventotto) provide a good definition
for energy-momentum and
angular momentum in $2+1$ dimensions; a positive energy theorem holds
with some characteristics which are intrinsic to $2+1$ dimensions.

\bs
{\bf 3. Solving Einstein's equations in 2+1 dimensions}

Gauges of geodesic type play a special role in $2+1$ dimensional
gravity. It was already pointed out in [\Moncrief] that in gaussian normal
coordinates the evolution equations in $2+1$ dimensions are reduced to
a system of ordinary differential equations. There is a variety of
geodesic gauges [\FerWal, \ModTol], the best known being the
Fermi-Walker gauge [\FerWal]. In the first order formalism such a
gauge is defined by
\def\treuno{3.1}
$$
\sum_i\xi^i\omega^a_{bi}=0~~~~
\sum_i\xi^i e^a_i=\sum_i\xi^i\delta^a_i.
\eqno(\treuno)
$$
These equations are solved by
\def\tredue{3.2}
$$
\omega^a_{bi}(\xi)=\xi^j\int^1_0 R^a_{bji}(\lambda \bfxi, t)\lambda d\lambda,
$$
$$
\omega^a_{b0}({\xi})=\omega^a_{b0}({\bf 0},t)+\xi^i\int^1_0
R^a_{bi0}(\lambda {\bfxi},t) d\lambda,
\eqno(\tredue)
$$
$$
e^a_i(\xi)=\delta^a_\mu+\xi^j\int^1_0\omega^a_{ji}(\lambda
{\bfxi},t)\lambda d\lambda
+\xi^j\int^1_0 S^a_{ji}(\lambda\bfxi,t) \lambda d\lambda.
$$
$$
e^a_0({\bf
\xi})=\delta^a_0+\xi^i\int^1_0\omega^a_{i0}(\lambda{\bfxi},t) d\lambda
+\xi^j\int^1_0 S^a_{j0}(\lambda\bfxi,t)d\lambda,
$$
where $R^a_{bji}$ and $S^a_{ji}$ are the Riemann tensor and the torsion.
Here we shall work with zero torsion. These resolvent formulas hold in
any dimensions, however in $2+1$
dimensions they assume a particular meaning because the Riemann tensor
appearing in eq.(\tredue) is given directly in terms of the energy-momentum.
Thus eqs.(\tredue) give the metric in terms of the sources by a
quadrature. However one has to keep in mind that the energy-momentum
tensor $T_c$ which appears in Einstein's equation
\def\tretre{3.3}
$$
\varepsilon_{abc} R^{ab}= -16\pi G T_c,\eqno(\tretre)
$$
\def\trequattro{3.4}
$$
R^{ab}=-8\pi G\varepsilon^{abc} T_c=-4\pi G
\varepsilon^{abc}~\varepsilon_{\rho\mu\nu}{\cal T}^{~\rho}_c dx^\mu\wedge
dx^\nu,
\eqno(\trequattro)
$$
is not completely arbitrary but is subject to
the symmetry and covariant conservation constraints summarized by
\def\trecinque{3.5}
$$
{\cal D} T^a=0,
{}~~~~
\varepsilon_{abc} T^b\wedge e^c =0.
\eqno(\trecinque)
$$
Thus the problem of solving Einstein's equations is reduced to that of
solving [\MS] the constraints (\trecinque). The conservation constraint is
automatically solved if we express the energy momentum tensor in terms
of the connection, through eq.(\tretre). The symmetry constraint is
more difficult. After introducing the
{}~~~cotangent ~~~vectors
$\displaystyle{T_{\mu}={\partial \xi^0\over\partial \xi^
\mu}}$,
$\displaystyle{P_{\mu}={\partial \rho\over\partial \xi^
\mu}}$ and
$\displaystyle{\Theta_{\mu}=\rho{\partial \theta\over\partial \xi^
\mu}}$ where $\rho$ and  $\theta$ are the polar variables in the
$(\xi^1,\xi^2)$ plane and writing the most general radial connection
in the form
\def\tresei{3.6}
$$
\omega^{ab}_\mu(\xi)=\varepsilon^{abc}\varepsilon_{\mu\rho\nu}P^\rho
A^\nu_c(\xi)
\eqno(\tresei)
$$
with
\def\tresette{3.7}
$$
A^\rho_c(\xi)= T_c \left [ \Theta^\rho\beta_1+T^\rho{(\beta_2-1)\over\rho}
\right ]+\Theta_c \left [\Theta^\rho\alpha_1 +T^\rho{\alpha_2\over\rho}
\right ]+P_c \left [\Theta^\rho \gamma_1+T^\rho{\gamma_2\over\rho}
\right ]
\eqno(\tresette)
$$
the metric becomes
\def\treotto{3.8}
$$
-ds^2= (A_1^2-B_1^2) dt^2 + 2 (A_1 A_2-B_1 B_2)dtd\theta + (A_2^2-B^2_2)
d\theta^2-d\rho^2
\eqno(\treotto)
$$
where $A_1+1$, $B_1$, $A_2$, $B_2$ are the primitives in $\rho$ of the
functions $\alpha_1$, $\beta_1$, $\alpha_2$, $\beta_2$. In terms of
such functions the symmetry
constraint becomes equivalent to the system [\MS]
\def\trenove{3.9}
$$
A_1\alpha_2^\prime -A_2 \alpha_1^\prime
+B_2\beta_1^\prime
 -B_1 \beta_2^\prime=0
$$
$$
\alpha_2 \gamma_1-\alpha_1 \gamma_2+A_2\gamma_1^\prime -A_1
\gamma_2^\prime
+{\partial\beta_1\over\partial\theta}-{\partial\beta_2\over \partial t}=0
\eqno(\trenove)
$$
$$
\beta_2 \gamma_1-\beta_1 \gamma_2+B_2\gamma_1^\prime -B_1
\gamma_2^\prime +{\partial\alpha_1\over\partial\theta}-
{\partial\alpha_2\over \partial t}=0.
$$
Three of the functions $\alpha_1$, $\beta_1$, $\gamma_1$, $\alpha_2$,
$\beta_2$, $\gamma_2$ can be given freely while the others are
determined by eq.(\trenove). We are interested in solutions in which the
source ${\cal T}^{a\rho}$ has some bounded support in space. The
condition for a
bounded support is that the invariants of the holonomies become
constant outside the sources. These conditions are very simple to
express in presence  of a Killing vector. For example if
$\displaystyle{\partial\over \partial \theta}$ is a Killing vector
they assume the form
\def\tredieci{3.10}
$$
\beta_2^2-\alpha_2^2-\gamma_2^2= {\rm const}
{}~~~~{\rm and}~~~~\alpha_2 B_2-\beta_2 A_2= {\rm const}
\eqno(\tredieci)
%$$
%and
%\def\treundici{3.11}
%$$
%
%\eqno(\treundici)
$$
while if $\displaystyle{\partial\over \partial t}$ is a Killing vector they
assume
the form [\MS]
\def\tredodici{3.11}
$$
\beta_1^2-\alpha_1^2-\gamma_1^2= {\rm const}
{}~~~~{\rm and}~~~~\alpha_1 B_1-\beta_1 A_1= {\rm const}.
\eqno(\tredodici)
$$
The first correspond to the conservation in time of the mass and
angular momentum, while the second correspond to the independence
outside the source of the Poincar\'e invariants of lines parallel to
the Killing vector $\displaystyle{\partial\over \partial t}$.

Both in presence of the Killing vector $\displaystyle{\partial\over
\partial \theta}$
and of the Killing vector $\displaystyle{\partial\over \partial t}$ the
constraint
equations can be solved by means of quadratures. For example in the
stationary  case we have [\MS]
\def\trequattordici{3.12}
$$
\alpha_2={B^2_1\over B^2_1-A^2_1}
{\partial\over \partial\rho}\left ({N\over  B_1}\right )+
2\alpha_1 I
$$
$$
\beta_2={A^2_1\over B^2_1-A^2_1}
{\partial\over \partial\rho}\left ({N\over A_1}\right )+
2\beta_1 I
\eqno(\trequattordici)
$$
$$
\gamma_2={B^2_1\over B^2_1-A^2_1}
{\partial\over\partial\theta}\left ({A_1\over B_1}\right )+
2\gamma_1 I ,
$$
where
\def\trequindici{3.13}
$$
I=\int^\rho_0 d\rho^\prime {N (A_1\beta_1-B_1 \alpha_1)\over (B_1^2-
A_1^2)^2}~;~ N={1\over 2\gamma_1}{\partial\over \partial \theta}
(A_1^2-B_1^2)
\eqno(\trequindici)
$$
By using these formulas or in the simpler cases by using directly
eqs.(\trenove) one can write down with great
ease all solution written in the literature [\Star] and also other
solutions [\MS].

In addition to the standard exterior ``Kerr'' solution [\Star, \DJtH]
\def\tresedici{3.14}
$$
ds^2=-(d t + 4 G J d\theta)^2+(1-4 G m)^2 (\rho - \rho_0)^2 d
\theta^2 + d\rho^2
\eqno(\tresedici)
$$
we can mention the exterior metric generated by a closed string with
tension [\Star]
\def\trediciassette{3.15}
$$
ds^2 = -k (\rho + c_1)^2(dt+4GJ d\theta)^2 +g_{\theta\theta} d\theta^2+ d\rho^2
\eqno(\trediciassette)
$$
with $k>0$ and $g_{\theta\theta}={\rm const} >0$
and the ``linear'' universes [\Star]
\def\trediciotto{3.16}
$$
ds^2=2 g_{t\theta}dt d\theta+ k (\rho + c_1) d\theta^2 + d\rho^2
\eqno(\trediciotto)
$$
\def\trediciannove{3.17}
$$
ds^2= -k(\rho + c_1) dt^2 + 2 g_{t\theta} dt d\theta^2 + d\rho^2
\eqno(\trediciannove)
$$
with $g_{t\theta}={\rm const}$. It is also possible [\MS] to write down time
dependent  solutions which satisfy the energy condition over all space-time.

\bs
{\bf Causality}

The discovery by Gott [\Gott] of a simple system in $2+1$ dimensions which
possesses closed time-like curves (CTC) has revived the issue of the
consistency of general relativity with causality [\Goedel,
\Tipler].
%The problem was
%raised by Einstein already in 1914 [\Einstein]
The first concrete example of a solution of
Einstein's equations which possesses CTC was given in 1949 by
Goedel [\Goedel]. Actually Goedel's example, due to the trivial dependence in
the
$z-$coordinate, is one of the first solution of $2+1$ dimensional
gravity (in presence of a negative cosmological constant). In Goedel's
universe the violation of causality occurs for very large trips. Also
the $2+1$ dimensional ``Kerr'' solution eq.(\dueventinove) as pointed
out in [\DJtH] and
easily checked, shows CTC near the origin, while at large
distances there are no CTC's. The problem of CTC's is in part related
to the energy theorem in the following sense: According to
the theorem [\CFGO,\MSener] which we discussed at the end of the last section,
in an open universe with time-like momentum there cannot be Gott
pairs. This decreases the probability that an open time-like universe
composed of point particles contains CTC's, but is not yet a proof of
their absence.

For a closed universe composed of a finite number of point-like
spinless particles 't Hooft [\tHuno] gave a general treatment which allows to
prove that even though the formation of Gott's pairs is energetically
permitted, the universe collapses before a traveler can complete a
CTC.

With regard to the occurrence of CTC's in the ``Kerr'' solution
eq.(\dueventinove)
such a problem was examined in [\MSctc] by employing the Fermi-Walker gauge
described in the previous section. The physical nature of such a gauge
allows to extract useful information from the energy condition which
here is used in the weak form (WEC) stating that for any time-like
$u^\mu$ we have $u^\mu T_{\mu\nu}u^\nu<0 $. The following
theorem holds [\MSctc]:

For a stationary open universe with axial symmetry if the
matter sources satisfy the WEC and there are no CTC at space infinity,
then  there are no CTC at all. Thus the ``singular  source'' related
to (\dueventinove)
%(\ref{kerr})
does not satisfy the WEC.

In fact from the WEC {\it and} the absence of CTC's at infinity it is
possible to prove that
$$
{d\over d\rho}({g_{\theta\theta}\over \det(e)}) > 0.
$$
As $g_{\theta\theta}=0$ for $\rho=0$ and  $\det(e)$ cannot vanish in
open universes with axial symmetry [\MSctc], $g_{\theta\theta}$ is always
positive. Thus the absence of CTC's is the result of a subtle interplay
between the WEC and boundary conditions.
The hypothesis of absence of CTC's at infinity is a necessary one as
one can construct examples of universes which satisfy the WEC but have
CTC's at space infinity [\MSctc].

With the same techniques the theorem on the absence of CTC's can also
be proved for all closed stationary universes with axial symmetry.
With regard to the extension to non axially symmetric stationary
universes at present the proof goes through provided $\det(e)$ in the
Fermi-Walker gauge never vanishes but one expects to be generally
true for stationary universes.

Due to the simple structure of $2+1$ dimensional gravity one would
expect a very general simple statement about CTC's
but at present this is lacking.

Finally we remark that in a model with point-like particles, such
particles have to be taken with zero angular momentum otherwise we
have a violation causality or equivalently of the energy condition.

%The general problem of CTC's in general relativity has been studied by
%Tipler and Hawking at the classical level and several other papers
%have appeared on the quantum aspects of the problem. At any rate a
%theory which supports CTC's at the classical level can hardly be
%considered as consistent unless we completely alter our concept of
%causality.
\bs
{\bf Canonical quantization}

The most direct approach to quantizing $2+1$ dimensional gravity is
the canonical approach. One starts fixing a priory the topology of
space-time. Compact 3-dimensional manifolds are to be avoided due to
causality requirements [\GerHor]. Then the simplest choice is ${\cal M}=
\Sigma\times R$ with $\Sigma$ a compact orientable two-dimensional
manifold. The
classical ADM [\ADM] hamiltonian formulation of gravity in $n+1$
dimensions starts with the
action
\def\quattrouno{4.1}
$$
A=\int (\pi^{ab}g_{ab,t}- N{\cal H}-N^a{\cal H}_a) dx^2 dt
\eqno(\quattrouno)
$$
with
\def\quattrodue{4.2}
$$
{\cal H}={1\over \sqrt{g}}[\pi^{ab}\pi_{ab}-(\pi^a_a)^2]-\sqrt{g}~R
\eqno(\quattrodue)
$$
and with
\def\quattrotre{4.3}
$$
{\cal H}_a = -2\nabla_b\pi^b_a
\eqno(\quattrotre)
$$
where $\nabla_b$ is the covariant derivative with respect to the space
metric of the $n$ dimensional space-like manifold defined by $t={\rm
const}, $ $g$ is the determinant of the metric of the space slice and
$R$ is the scalar curvature of the space slice. It is well known that the role
of the lapse and shift
variables $N$ and $N^a$ is that of Lagrange multipliers. We know that for
$\Sigma\approx S^2$ the connections are trivial and thus the model is
trivial. There are two
ways to give the theory a non trivial content; either one introduces a non
trivial space topology or one adds matter. The first alternative is the
simplest. Thus one takes for $\Sigma$ an orientable two dimensional
manifold of genus $g\ge 1$.

Moncrief [\Moncrief] and Hosoya and Nakano [\HosNak] have shown that
the classical system described by (\quattrouno) in $2+1$
dimensions is equivalent to a hamiltonian system of $6g-6$ degrees of
freedom for $g >1$ and $2$ degrees of freedom for $g=1$ i.e. for the
torus. The reductions to a finite number of degrees of freedom is not
surprising if one takes into account that in $2+1$ dimensions there
are no gravitons and thus all the dynamical role is played by the non
trivial independent holonomies of $\Sigma$. The method is to introduce
a foliation of space-time with space-like surfaces of constant
exterior curvature $K$. Such a quantity plays the role of
time. The metric description of the space-like surface of constant $K$
is provided by  a metric $h_{ab}$ of constant scalar curvature ( $-1$ for $g>1$
and $0$ for $g=1$) and a conformal factor $e^{2\lambda}$
\def\quattroquattro{4.4}
$$
g_{ab}=e^{2\lambda} h_{ab};~~~~ R(h)=-1~~~{\rm or}~~~0~~ {\rm for}~~~g=1
\eqno(\quattroquattro)
$$
plus the Teichmueller parameters $\tau_\alpha$ which are 2 for $g=1$
and $6g-6$ for $g>1$. The momentum constraints
${\cal H}_a=0$
impose on the conjugate momenta the structure $\pi^{ab}=\pi^{abTT}
+{1\over 2} K\sqrt{g}~ g^{ab}$ with $\pi^{TT}$ transverse and
traceless,
while the hamiltonian constraint ${\cal H}=0$ determines the conformal
factor $\lambda$ as the solution of a non linear elliptic differential
equation. There are [\Moncrief] existence and uniqueness theorems for the
solution of such differential equation. At this stage is not difficult to
perform the reduction of the degrees of freedom. One substitutes in
(\quattrouno )
${\cal H}_a=0$, ${\cal H}=0$  and defined
\def\quattrocinque{4.5}
$$
p_\alpha=\int_\Sigma e^{2 \lambda} \pi^{abTT}(x) {\partial h_{ab}\over
\partial \tau_\alpha}(x,\tau)dx^2
\eqno(\quattrocinque)
$$
one reaches apart from some boundary terms which are not relevant for
the equations of motion the action
\def\quattrosei{4.6}
$$
A=\int [p_\alpha {d\tau_\alpha\over dt} -{dK\over dt}\int_\Sigma
\sqrt{g}~ dx^2] dt.
\eqno(\quattrosei)
$$
Thus the Hamiltonian is simply given by the area of the space-like
slice $K={\rm const}$; however in the canonical formalism one has to
express such an area in terms of the canonical variables. In the
simple case of the torus this can be accomplished with the result
[\Moncrief , \Carlip]
\def\quattrosette{4.7}
$$
\int_\Sigma \sqrt{g}~ dx^2= {1\over K}[\tau^2_2(p_1^2+p^2_2)]^{1/2}.
\eqno(\quattrosette)
$$
The $K$ dependence in the Hamiltonian can ~be ~removed [\Moncrief] ~by
{}~putting $K=\exp(t/(2\pi)^2)$.
One can now proceed to the canonical quantization by posing
$\displaystyle{p_j=-i{\partial\over \partial\tau_j}}$ thus reaching the
Schroedinger equation
\def\quattrootto{4.8}
$$
i{\partial\psi(\tau,t)\over \partial t}=[-\tau_2^2 \nabla^2]^{1/2}\psi(\tau,t).
\eqno(\quattrootto)
$$
For higher genus the expression of the reduced Hamiltonian is given
only implicitly through the solution of a non
linear elliptic differential equation.

The eigenvalue equation associated to eq.(\quattrootto) is of the
Bessel type and
thus appears to be easily soluble but this is not so, because one
should impose on the solution the invariance under modular
transformations of $\tau=\tau_1+i\tau_2$
\def\quattronove{4.9}
$$
\tau\rightarrow \tau+1,~~~~\tau\rightarrow -{1\over \tau}
\eqno(\quattronove)
$$
which correspond to the large
diffeomorphysms.

The ADM approach is the typical second order approach to gravity. One
can however formulate gravity also in the first order approach with
dreibeins and connections taken as independent variables and which
shares a greater similarity with the usual gauge theories. The
action
%$$
%A=\int R^{a_1 a_2}\wedge e^{c}\dots e^{a_D}\varepsilon_{a_1\dots a_D}
%$$
in $2+1$ dimensions is given by
\def\quattronoveb{4.10}
$$
A=-{1\over 2}\int R^{a b}\wedge e^{c} ~\varepsilon_{abc}
\eqno(\quattronoveb)
$$
which apart a divergence term has, in the notation employed in sect.3,
the  form
\def\quattrodieci{4.11}
$$
A=\int dt\int dx^2 \epsilon^{ij} (-\omega_{aj}\dot e^a_i + e^a_0
\epsilon_{abc}R^{bc}_{ij} + \omega_{a0}S^a_{ij})
\eqno(\quattrodieci)
$$
where $R^{ab}$ is the curvature and $S^a$ is the torsion.
{}From eq.(\quattrodieci) one derives the canonical P.B.
\def\quattroundici{4.12}
$$
\{e^a_i(x),\omega^a_j(y) \}=-\epsilon_{ij}\delta^2(x-y)
\eqno(\quattroundici)
$$
and two constraints $R^{ab}\approx 0$ and $S^a\approx 0$.
Thus we have that the group curvature of
the Poincar\'e connection is locally zero and the only non trivial
quantities are the holonomies related to non contractible loops. It is
of great interest to study the algebra of such holonomies or better of
the invariant quantities which characterize them. We recall that there are
two invariants for each holonomy i.e. the ``angle'' of the Lorentz
transformation
and the projection of the ``displacement vector'' on the ``rotation
axis''. An equivalent choice of invariants is given by $q=\tr {\cal M}$
and $\nu={\cal M}^a_{~b}\varepsilon_{a~c}^{~b} E^c$, where ${\cal M}$ is
the Lorentz transformation in the adjoint representation.
Starting from the canonical
P.B. eq.(\quattroundici) Nelson and Regge [\NelReg, \NelRegdue] have derived
the P.B. of $q$ and $\nu$
relative to two loops $u$ and $v$ with a single intersection. They are
[\NelReg]
\def\quattrododici{4.13}
$$
\{q(u),\nu(v)\}={1\over 2}(q(uv)-q(uv^{-1}))
\eqno(\quattrododici)
$$
and
\def\quattrotredici{4.14}
$$
\{\nu(u),\nu(v)\}={1\over 2}(\nu(uv)-\nu(uv^{-1}))
\eqno(\quattrotredici)
$$
all the other P.B. as well as those of non intersecting loops being zero.
Using such relations one can compute recursively for example, all the
invariants of the loops generated by two intersecting loops. We know
however that there are only a finite number of invariant quantities (
$4$ for the torus and $12g-12$ for $g>1$) and thus we have the problem
of the reduction to a minimal set of invariant and that of giving the
representation of this algebra at the quantum level.

For the torus we have only two generators for the group $\pi_1$ and in
this simple case the program can be easily completed with the
following result. By means of a global Poincar\'e transformation one
can reduce both holonomies related to the two generators $u$ and $v$ to
the form of a boost in the $t,x$ plane and a translation in the $y$
direction. If we call $\lambda$ the boost parameter and with
$a$ the translation for the $u$ loop and $\mu$ and $b$ those for the
$v$ loop we have
from eqs.(\quattrododici,~ \quattrotredici) the following result
\def\quattroquattordici{4.15}
$$
\{\lambda,2b\}=\{\mu,-2a\}=1
\eqno(\quattroquattordici)
$$
all the other P.B. being zero.

The same program has been carried through by Nelson and Regge
explicitly for $g=2$ and implicitly for any $g$ [\NelRegdue]. It is of interest
to relate this approach which contains only constants of motion to the
ADM-type of approach described previously. Clearly some space-like
foliation has to be performed to introduce the concept of time. This
has been accomplished for the torus by Carlip [\Carlip, \Carlipdue]
by introducing the
foliation $t=K^{-1} \cosh(\xi)$  and $x=K^{-1} \sinh(\xi)$ , $K$ constant
and $\xi$ and $y$ variable, where $K$ is just the extrinsic curvature
of the two dimensional surface embedded in Minkowski space and plays
the role of time as in the Moncrief approach.

It is simple to compute the complex modulus of the space-slice
\def\quattroquindici{4.16}
$$
\tau=(a + i K^{-1} \lambda)^{-1}(b + i K^{-1} \mu)
\eqno(\quattroquindici)
$$
while the area and thus the Hamiltonian (see eq.(\quattrosei))
becomes [\Carlip]
\def\quattrosedici{4.17}
$$
H={a\mu-\lambda b\over K}.
\eqno(\quattrosedici)
$$
Again the wave function will be required to be invariant under modular
transformations expressed in the new $\lambda, \mu$ coordinates.
One can then relate the two approaches i.e. change from the $\lambda,
\mu$ representation to the $\tau_1,\tau_2$ representation through the
following steps [\Carlipdue]: a) Compute the $p_1$, $p_2$ conjugate to
$\tau_1$, $\tau_2$ as a function of the variables $\lambda
,a$. b) Compute the eigenfunctions
$\psi_\tau(\lambda,\mu)$ of $\tau$. Being $\tau$
eq.(\quattroquindici) a normal operator the eigenfunctions of $\tau$
are simultaneously eigenfunctions of $\tau_1$ and $\tau_2$.
c) Compute the $p_1$, $p_2$ in terms of
$\tau_1$, $\tau_2$. The result is [\Carlip]
\def\quattrodiciassette{4.18}
$$
p_1=-i{\partial\over \partial \tau_1};~~~~p_2=-i{\partial\over
\partial \tau_2}+{i\over \tau_2}
\eqno(\quattrodiciassette)
$$
and substituting into eq.(\quattrosedici) one yields for the Hamiltonian
\def\quattrodiciotto{4.19}
$$
H^2=K^{-2}(-\tau_2^2(({\partial\over \partial \tau_1})^2+({\partial\over
\partial \tau_2})^2) + i\tau_2 {\partial\over \partial \tau_1} -{1\over 4}).
\eqno(\quattrodiciotto)
$$
The request of invariance under
modular transformations in the $\lambda, \mu$ space imposes the
eigenfunction of eq.(\quattrodiciotto) to transform like modular form
of weight $1/2$.
Thus  while the two approaches are equivalent at the classical level
they are not at the quantum level.
In a series of papers to which we refer for full details, Carlip
[\Carlipdue] and Carlip and Nelson [\CarNel] have
analyzed the source of this difference. The use of one or the other
approach leads to a natural choice of fundamental variables which
differ in the two cases. The translation of the classical theory to
the quantum theory, as is
well known, is subject to the problem of the ordering  of the
operators, which at least in part is related to the choice of the
metric defining the scalar product in Hilbert space.

\bs
{\bf 5. Tessellation and polygonal approaches}

Usually the lattice is employed in field theory as a non perturbative
regulator. In $2+1$ dimensional gravity due to the local flatness of
space-time it is possible to set up lattice-like schemes which are
exact.

First we describe the tessellation approach of 't Hooft [\tHuno,\tHdue].
The main idea
is to give a foliation of space-time in terms of two dimensional
space-like surfaces which are piecewise flat and thus are built up by
gluing together polygons. To have the time variable to flow at the
same rate on two adjacent polygons one must choose the Lorentz frame
on each polygon in such a way that the seam between the two polygons
moves with the same (and ``opposite'') speed with respect to the two
Lorentz frames. The geometry of a given time slice is described by the
length of the sides $L_i$ and the angles $\alpha_i$ of the polygons and by the
magnitude $2\eta_i$ of the Lorentz boost across the side $i$. Actually
it is possible to compute the angles of the polygons from the boosts
$\eta_i$ by imposing the compatibility of the Lorentz transformations
across the sides which converge to a single vertex. Thus the $\eta_i$
and $L_i$ can be chosen as the fundamental variables in this approach.

There are constraints on the fundamental variables $L_i$ and
$\eta_i$ because to close each polygon we must have
\def\cinqueuno{5.1}
$$
\sum_i( \pi-\alpha_i)=2\pi
\eqno(\cinqueuno)
$$
and
\def\cinquedue{5.2}
$$
\sum_i e^{i \alpha_i}L_i=0
\eqno(\cinquedue)
$$
where the sum extends to each single polygon.

Particles can be introduced by removing from a polygon a sector whose
opening is given by $\delta=8\pi G m $ if the particle is at rest with
respect to the Lorentz frame of the polygon. If the particle moves
with respect to such frame a proper boost has to be applied. Due to
the presence of the boosts and the motion of the particles, the lengths
$L_i$ of the sides change in time and can also vanish. Other
sides can also originate for the same reason and these processes are
encoded in nine transition rules which are described in detail in
[\tHdue]. Such a deterministic system has been employed by 't Hooft to prove
the absence of formation of CTC's in a closed universe containing
point-like spinless particles  and to study by means of numerical
calculations simple three dimensional cosmological models. To proceed to
quantization  one has first to set up an hamiltonian description of
the system. There are two elegant features in the hamiltonian
formalism: the first are the remarkably simple P.B.
\def\cinquetre{5.3}
$$
\{2 \eta_i,L_j\}=\delta_{ij}
\eqno(\cinquetre)
$$
and the second the
fact that the Hamiltonian is given by all the deficit angles
\def\cinquequattro{5.4}
$$
H= {1\over 8 \pi G}\sum_V (2\pi - \sum_i \alpha_i(V)).
\eqno(\cinquequattro)
$$
These deficits are both due
to the presence of a particle and to the fact that $\sum_i\alpha_i(V)$
extended to the angles converging to a single vertex $V$ is not $2
\pi$ due to the presence of the boosts. Recalling that the
$\alpha_i(V)$ can be computed from the boosts we have that the
Hamiltonian is only  function of the boosts. One has
however keep in mind that in addition to the constraints
eq.(\cinqueuno, \cinquedue) there
are other constraints of the type of triangular inequalities among the
boosts which have to be taken into account [\tHdue]. Quantization is achieved
by replacing the P.B. with commutators and the classical transition rules
mentioned above now play the role of boundary conditions on the wave
function. A remarkable consequence of the form (\cinquequattro) of the
Hamiltonian is that as the $\alpha_i$ are determined in terms of the
$\eta_i$ by inverting trigonometric relations, the $\alpha_i$ and as
a consequence $8 \pi  G H$ is determined only modulo $2\pi$.
This means that the evolution operator $\exp( -iHt)$ is well defined
only for discrete values of the time parameter $t= 8 \pi G n$.
It would be nice to apply this quantization scheme to some simple
instances.

Waelbroeck's polygonal approach [\Wael] is strictly related to the first
order formulation  of gravity and to the algebra of observables of
Nelson and Regge of which it can be considered as a geometrical
realization. The space-like slice is described by a
polygon in flat
Minkowski space in which pairs of sides are identified. Thus we have
as variables the three vectors representing the sides and for each
side an element of the Lorentz group which represents the Lorentz holonomy
which takes from the given side to its image.
One denotes the sides by $E(\mu)$ with $\mu=1,\dots 2g$ with $g$ the
genus of the surface and the Lorentz holonomies which relate the side
$\mu$ with its image by $M(\mu)$. Thus the Lorentz component of the
holonomy related to a closed path which joins $E(\mu)$ with its image
is simply given by $M(\mu)$ and it is also possible to reconstruct the
translation part $E_\sigma$ of the Poincar\`e holonomy related to a
closed loop $\sigma$, in terms of
the $M(\mu)$
and the $E(\mu)$ [\Wael].

There are two constraints: the first is that the polygon must close i.e.
\def\cinquequattrob{5.5}
$$
\sum_{\mu=1}^{2g}(I-M^{-1}(\mu))E(\mu)=0
\eqno(\cinquequattrob)
$$
and the other, like in the algebra of observables, tells us that the
Lorentz relator equals the identity
\def\cinquecinque{5.6}
$$
W = M(1)M^{-1}(2)M(1)^{-1}M(2)\dots M(2g-1)M^{-1}(2g)M(2g-1)^{-1}M(2g)=I.
\eqno(\cinquecinque)
$$
%Each of the constraints gives rise to three first class constraints.

Spinless point particles can be introduced by inserting along the
polygon two consecutive sides, one given by $E(\mu)$, $\mu=2g+1\dots
2g+N$, $N$ being the number of particles, and the other by its image
$M^{-1}(\mu)E(\mu)$ under the time-like
Lorentz transformation $M(\mu)$ subject to the constraint that its
rotation angle is
related to the mass of the particle by $\delta(\mu) = 8\pi G m(\mu)$ or
equivalently $\tr M(\mu)=2 \cos^2\delta(\mu) + 1$.
The closure condition of the polygon has to be accordingly modified to
include these additional couples of sides. As in the ADM approach the
Hamiltonian  is provided by a combination of the constraints with
lapse and shift functions
\def\cinquesei{5.7}
$$
H=\sum_a {1\over 2}N_a \varepsilon^{a~c}_{~b} W^b_{~c} +
\sum_{\mu=2g+1}^{2g+N} N(\mu)~ \tr M(\mu),
\eqno(\cinquesei)
$$
the second sum representing the contribution of the particles. The
three-vector $N_a$ is a time-like vector which can be represented by a
vector placed e.g. at the origin of the side $E(1)$ and by its
images obtained by parallel transport along the polygon. The $N(\mu)$
are in principle arbitrary constants.
The correct equations of motion are generated by the following
simplectic structure
\def\cinquesette{5.8}
$$
\{E^a(\mu), E^b(\mu)\}=\varepsilon^{ab}_{~~c}E^c(\mu)
\eqno(\cinquesette)
$$
\def\cinqueotto{5.9}
$$
\{E^a(\mu), M^b_{~c}(\mu)\}=\varepsilon^{ab}_{~~d} M^d_{~c}(\mu)
\eqno(\cinqueotto)
$$
all the others P.B. being 0, combined with the Hamiltonian $H$, and
they are
\def\cinquenove{5.10}
$$
{d M(\mu)\over dt}=0
\eqno(\cinquenove)
$$
\def\cinquedieci{5.11}
$$
{d^2 E(\mu)\over dt^2}=0
\eqno(\cinquedieci)
$$
which simply express that the Lorentz holonomies are constants of
motion and that the sides $E(\mu)$ (with this choice of time) vary
linearly in time. The constraints (\cinquequattrob, \cinquecinque) are
first order constraints. In order to proceed to quantization one has to give
a representation of the eqs.(\cinquesette, \cinqueotto) in terms of
coordinates and conjugate momenta. Defining the vector
\def\cinqueundici{5.12}
$$
P_a(\mu)={1\over 2}\varepsilon^{~b}_{a~c} M^c_{~b}(\mu)
\eqno(\cinqueundici)
$$
the conjugate vector to it is [\Wael]
\def\cinquedodici{5.13}
$$
X(\mu)={1\over P^2(\mu)}(P(\mu)\wedge J(\mu)-
{2 (E(\mu)\cdot P(\mu)) P(\mu)\over \tr M(\mu)-1})
\eqno(\cinquedodici)
$$
with $J(\mu)\equiv(1-M^{-1}(\mu)) E(\mu)$.
Thus one can pose
\def\cinquetredici{5.14}
$$
[X^a(\mu),P_b(\mu)]=i\delta^a_b
\eqno(\cinquetredici)
$$
and then represent the momentum as $\displaystyle{P_b(\mu)=-i\partial /
\partial X^b(\mu)}$. The relation between the fundamental variables
$M^a_{~b}(\mu)$ and $E^a(\mu)$ and the $X^a(\mu)$ and $P_a(\mu)$ are
algebraic and even though invertible they contain square roots in non
simple combinations.
The simplest quantum mechanical application is the writing of the
equations corresponding to the constraints and to the Wheeler-de Witt
equations. The Lorentz constraint equations are
\def\cinquequattordici{5.15}
$$
J^a\psi(X)=0
\eqno(\cinquequattordici)
$$
where $J^a$ in the new variables $X$ and $P$ takes the simple form
\def\cinquequindici{5.16}
$$
J=\sum_\mu X(\mu)\wedge P(\mu)
\eqno(\cinquequindici)
$$
and the Wheeler de Witt equations are
\def\cinquesedici{5.17}
$$
P^a\psi(X)=0.
\eqno(\cinquesedici)
$$
Eq.(\cinquequattordici) simply tells us that $\psi$ has to be a scalar
in the  $X(\mu)$
variables. We recall that $P_a= \varepsilon_{a~c}^{~b} W^c_{~b}/2$ and $W$
is a function of the $P_a(\mu)$ containing square roots of
$P^2(\mu)\pm 1$ and thus is a non rational function of
$\displaystyle{-i\partial\over \partial X^a(\mu)}$. This is similar to
what is found in the ADM and first order formulation of quantum
gravity already in the simple instance of the torus in absence of
particles.

An alternative to writing the Wheeler de Witt equation is to introduce
an internal time, like the length of a side of the polygon and by
using the constraints solve the conjugate momentum to that variable in
terms of the remaining variables. That gives the Hamiltonian relative
to the given choice of internal time and one can proceed to write down
the Schroedinger equation. Such an approach however appears to be
still inequivalent to the previously described approaches [\Wael].

\bs
{\bf Conclusions}

We saw that $2+1$ dimensional gravity is well understood at the
classical level even if there are still some problems to be solved. At the
quantum level on the other hand we have a variety of approaches all
of which appear to lead to inequivalent formulations. In addition only
the simplest situations, e.g. absence of particles, and the simplest
topologies have been thoroughly examined. One has to sort out which
approach is the correct one and possibly examine more general situations.

\bs

{\bf Acknowledgments}

I am grateful to the organizers of the conference for the kind
invitation and to E. Guadagnini, R. Loll  and especially D. Seminara
for useful discussions.
\vfill
\eject

{\bf References}

[\Star] Staruszkiewicz A., Acta. Phys. Polon. {\bf 24} (1963) 734; Leutwyler
H., Nuovo Cimento {\bf 42A} (1966) 159;
Deser S., Jackiw R. and  't Hooft G., Ann. Phys.
(N.Y.)  {\bf 152} (1984) 220; Gott J.R. and
Alpert M.,   Gen. Rel. Grav. {\bf 16} (1984) 243;
Giddings S., Abbot J. and Kuchar K.,  Gen. Rel.Grav. {\bf 16} (1984) 751;
Clement G., Int. J. Theor. Phys. {\bf 24} (1985) 267;
Ann. Phys. (N.Y.) {\bf 201}
(1990) 241;  Grignani G. and Lee C.,  Ann. Phys. (N.Y.) {\bf 196} (1989) 386.

[\Vile] For a review on the subject see Vilenkin, A., Phys.Rep. {\bf C 121}
(1985) 263.

[\Callum] Levi-Civita T., Rend.Accad.Lincei {\bf 26} (1917) 307;
Kramers D., Stephani H., Herlt E. and MacCallum M., {\it ``Exact
solutions of Einstein's field equations''}, Cambridge University Press,
Chapt.20 .

[\Wein] Weinberg S., {\it ``Gravitation and cosmology''} J. Wiley \& Sons;
Nester J.,Phys.Lett. {\bf 83A} (1981) 241.

[\DJtH] Deser S., Jackiw R. and `t Hooft G.,
Ann. Phys.(N.Y.) {\bf 152} (1984) 220.

[\NelReg] Nelson J.E. and Regge T., Nucl.Phys. {\bf B328} (1989) 190.

[\AchTow] Achucarro, A. and Townsend, P.K., Phys.Lett. {\bf B180} (1986) 89;
Phys.Lett. {\bf B229} (1989) 383.

[\Witten] Witten E., Nucl.Phys. {\bf B311} (1989) 46.

[\Scho] Schoen  P., Yau  S.T., Comm.Math.Phys. {\bf 65} (1979) 45; Witten
E., Comm.Math.Phys. {\bf 80} (1981) 381; Parker T. and Taubes C.H.,
Comm.Math.Phys. {\bf 84} (1982) 221; Nester J.,Phys.Lett. {\bf 83A}
(1981) 241.

[\MSener] Menotti P. and Seminara D., {\it ``Energy theorem for 2+1
dimensional gravity''} Preprint MIT-CTP \# 234 IFUP-TH-33/94.

[\HawEll] Hawking S.H. and Ellis G.F.R.,{\it `` The large scale structure of
the universe''}, Cambridge University Press (1973).

[\Gott] Gott J.R., Phys. Rev. Lett.  {\bf 66} (1991) 1126.

[\Cut] Deser S., Jackiw R. and 't Hooft G., Phys.Rev.Lett. {\bf 68}
(1992) 2647; Ori A., Phys.Rev. {\bf D44} (1992) R2214; Cutler C., Phys.Rev.{\bf
D45} (1992) 487; Kabat D., Phys. Rev. {\bf D46} (1992) 2720.

[\CFGO] Carroll S. M., Farhi E., Guth A. H. and Olum K. O.,
{it `` Energy-momentum tensor restrictions on the creation of Gott-time
machine''}, CTP \# 2252, gr-qc/9404065.

[\Moncrief] Moncrief V., J.Math.Phys. {\bf 30}  (1989) 2907.
%Class.Quant.Grav. 7(1990)163;

[\FerWal]  Fermi E., Atti Accad. Naz. Lincei Cl. Sci. Fis. Mat. \&
Nat. {\bf 31} (1922) 184, 306; Walker A.G., Proc. London Math. Soc.
{\bf 42} (1932) 90;

[\ModTol] Modanese G. and Toller M., J. Math. Phys. {\bf 31}
(1990)452.

[\MS] Menotti P. and Seminara D., Ann. Phys. (N.Y.) {\bf 208} (1991) 449;
Nucl.Phys. {\bf B376} (1992) 411.

%[\Einstein] Einstein A.,

[\Goedel] Goedel K., Rev.Mod.Phys. {\bf 21} (1949) 447.

[\Tipler] Tipler F.J., Ann.Phys. (NY) {\bf 108} (1977) 1; Hawking S.W.,
{\it Phys.Rev. D } {\bf 46} (1992) 603

[\tHuno] 't Hooft G., Class. Quantum Grav. {\bf 9} (1992) 1335.

[\tHdue] 't Hooft G., Class. Quantum Grav. {\bf 10} (1993) 1023.

[\MSctc] Menotti P. and Seminara D., Phys. Lett. {\bf B301} 25;
Nucl.Phys. {\bf B419} (1994) 189.

%[\Soleng] Soleng H., Phys.Rev. {\bf D49} (1994) 1124

[\GerHor] Geroch R. and Horowitz G.T., {\it ``Global structure of
spacetimes''} in
{\it `` General relativity: An Einstein centenary survey''}. S.W. Hawking and
W. Israel eds. Cambridge University Press.

[\ADM] Arnovitt R., Deser R. and Misner C. J.Math.Phys. {\bf 1} (1962) 434;
Nuovo Cimento {\bf 19} (1961) 668; in {\it ``Gravitation: An Introduction to
Current Research''}, L. Witten ed. Jhon Wiley and Sons, New York.

[\HosNak] Hosoya A. and  Nakao K., Class.Quantum Grav. {\bf 7} (1990) 163;
Progr.Theor.Phys. {\bf 84} (1990) 739.

[\NelRegdue] Nelson J.E. and Regge T.,Phys. Lett. {\bf B273} (1991) 213;
Comm.Math.Phys. {\bf 155} (1993) 561;
Comm.Math.Phys. {\bf 141} (1993) 211;
Nelson J.E., Regge T. and Zertuche F., Nucl.Phys. {\bf B339} (1990)
516; for the choice of independent invariants in configuration space
see Loll R., {\it ``Independent Loo Invariants for 2+1 Gravity''}
gr-qc-9408007.

[\Carlip] Carlip S., {\it Phys.Rev.} {\bf D42} (1990) 2647

[\Carlipdue] Carlip S., {\it Phys.Rev.} {\bf D45} (1992) 3584;
{\it Phys.Rev.} {\bf D47} (1993) 4520;
{\it Class.Quantum Gravity} {\bf 8} (1991) 5;
{\it ``Six ways to quantize (2+1)-dimensional gravity''}
preprint UCD-93-15, NSF-ITP-93-63 gr-qc/9305020 (1993);
{\it Time (2+1)-dimensional gravity}
preprint UCD-94-18, gr-qc/9405043 (1993).

[\CarNel] Carlip S. and Nelson J.E., Phys.Lett. {\bf B324} (1994) 299;
``Comparative quantizations of 2+1 dimensional gravity''  UCD-94-37,
gr-qc 9411031.

[\Wael] Waelbroeck H., Nucl.Phys. {\bf B364} (1991) 475;
Phys.Rev. {\bf D50} (1994) 4982;
Waelbroeck H. and Zertuche F., Phys.Rev. {\bf D50} (1994) 4966.

\bye